\documentclass[a4paper,twoside,10pt]{letter}
\usepackage{graphicx,saj,multicol,subeqnarray}

\def\papertype{}

\begin{document}
\baselineskip=3.1truemm
\columnsep=.5truecm
\newenvironment{lefteqnarray}{\arraycolsep=0pt\begin{eqnarray}}
{\end{eqnarray}\protect\aftergroup\ignorespaces}
\newenvironment{lefteqnarray*}{\arraycolsep=0pt\begin{eqnarray*}}
{\end{eqnarray*}\protect\aftergroup\ignorespaces}
\newenvironment{leftsubeqnarray}{\arraycolsep=0pt\begin{subeqnarray}}
{\end{subeqnarray}\protect\aftergroup\ignorespaces}
\newcommand{\D}{$^\circ$}
\newcommand{\HI}{H\,{\sc i}}
\newcommand{\NaI}{Na\,{\sc i}}
\newcommand{\HeI}{He\,{\sc i}}
\newcommand{\HeII}{He\,{\sc ii}}
\newcommand{\FeVII}{[Fe\,{\sc vii}]}
\newcommand{\FeI}{Fe\,{\sc i}}
\newcommand{\FeII}{[Fe\,{\sc ii}]}
\newcommand{\FeIII}{[Fe\,{\sc iii}]}
\newcommand{\HII}{H\,{\sc ii}}
\newcommand{\CaII}{Ca\,{\sc ii}}
\newcommand{\hii}{h\,{\sc ii}}
\newcommand{\SII}{[S\,{\sc ii}]}
\newcommand{\SVI}{[S\,{\sc vi}]}
\newcommand{\SIII}{[S\,{\sc iii}]}
\def\p0{\phantom{0}}
\newcommand{\tbsp}{\rule{0pt}{10pt}}
\def\arcmin{\hbox{$^{\prime}$}}
\def\arcsec{\hbox{$^{\prime\prime}$}}

\newcommand{\OI}{[O\,{\sc i}]}
\newcommand{\OII}{[O\,{\sc ii}]}
\newcommand{\OIII}{[O\,{\sc iii}]}
\newcommand{\NII}{[N\,{\sc ii}]}
\newcommand{\NeIII}{[Ne\,{\sc iii}]}
\newcommand{\ArIII}{[Ar\,{\sc iii}]}



\markboth{\eightrm Optical Spectra of Radio Planetary Nebulae in the Small Magellanic Cloud}
{\eightrm J. L. Payne, M. D. Filipovi\'c, E. J. Crawford, A. Y.~De~Horta, G. L. White and F. H. Stootman}

{\ }

\publ

\type

{\ }


\title{Optical Spectra of Radio Planetary Nebulae in the Small Magellanic Cloud}


\authors{J. L. Payne$^{1}$, M. D. Filipovi\'c$^{2}$, E. J. Crawford$^{2}$, A.Y.~De~Horta$^{2}$, G. L. White$^{1}$ and F.H.~Stootman$^{2}$}

\vskip3mm

\address{$^1$Centre for Astronomy, James Cook University\break Townsville QLD, 4811, Australia}

\address{$^2$University of Western Sydney, Locked Bag 1797\break Penrith South, DC, NSW 1797, Australia}


\dates{Feburary 15, 2008}{TBA, 2008}


\summary{We present preliminary results from spectral observations of four (4) candidate radio sources co-identified with known planetary nebulae (PNe) in the Small Magellanic Cloud (SMC). These were made using the Radcliffe 1.9-meter telescope in Sutherland, South Africa. These radio PNe were originally found in Australia Telescope Compact Array (ATCA) surveys of the SMC at 1.42 and 2.37~GHz, and were further confirmed by new high resolution ATCA images at 6 and 3 cm (4\arcsec/2\arcsec). Optical PNe and radio candidates are within 2\arcsec\ and may represent a sub-population of selected radio bright objects.  Nebular ionized masses of these objects may be 2.6~$M_\odot$ or greater, supporting the existence of PNe progenitor central stars with masses up to 8 $M_\odot$.} 



\begin{multicols}{2}
{


\section{1. INTRODUCTION}

Ninety five percent of all stars will end their lives as white dwarfs.  Planetary nebulae (PNe) represent a relatively short ($10^{4}$ yr) phase between the asymptotic giant branch (AGB) and white dwarf, having complex morphological features such as differences in stellar wind properties over time. These magnificent objects possess ionised, neutral atomic, molecular and solid states of matter in diverse regions having physical environments that range in temperature from $10^{2}$ K to greater than $10^{6}$ K with an average electron density of  $10^{3}$ cm$^{-3}$.  Although these objects radiate from the \mbox{X-ray} to the radio, detected structures are influenced by selection effects from intervening dust and gas, instrument sensitivity and distance (see Kwok (2005) for a more detailed discussion).

Most PNe have central-star and nebular masses of only about 0.6 and 0.3 $M_\odot$, respectively. However, detection of white dwarfs in open clusters suggests the main-sequence mass of PNe progenitors can be as high as 8 $M_\odot$ (Kwok 1994).

Optical spectral requirements for the confirmation of PNe have been summarised very nicely by Reid \& Parker (2006). Classically, PNe can be identified by a \OIII 5007\AA :\OIII 4959\AA:H$\beta$  intensity ratio near 9:3:1.  However, this can be relaxed when \NII  6583\AA\ is greater than H$\alpha$.  In that case, a strong \OIII\ line has been often detected along with the high excitation \HeII  4686\AA\ line hardly seen in \HII\ regions.

Generally, the \OII  3727\AA\ doublet is seen in PNe, as well as \NeIII  3869\AA, \ArIII  7135\AA\ and \HeI  6678\AA\ lines. \SII  6717, 6731\AA\  is usually present but
not significant when compared to H$\alpha$.

In this paper we present optical spectra of four (4) PNe that co-identify with radio PNe candidates recently found in the Small Magellanic Cloud (SMC). Section 2 details our spectral observations and reduction methods used to calibrate the data analysed  in Section 3. The paper concludes with some final comments and a brief summary as presented in Section~4.

\section{2. OBSERVATIONS}\label{observations}

Beginning with SMC J004336-730227 and SMC J005730-723224, as described in Payne et al. (2004), 28 candidate radio PNe have been found within a few arcseconds of known optical PNe in the direction of the
Magellanic Clouds (MCs; Filipovi\'c et al., in preparation). This includes six (6) SMC sources from
Australia Telescope Compact Array (ATCA) observations, completed for short-spacing using Parkes data and
the {\sc miriad} software suite to create mosaic images at 1.42 and 2.37~GHz (see Filipovi\'c et al. 2002).

Recent followup ATCA observations at 6 and 3 cm give much higher resolutions (4\arcsec and 2\arcsec, respectively) and appear to confirm these objects as bright radio counterparts to within 2\arcsec\ of known optical PNe. The number and flux densities (up to ten times greater than expected) of these candidate
radio PNe are unexpected given the distance of the SMC of $\sim60$ kpc (Hilditch et al. 2005).  This may well modify our current understanding of PNe, including their progenitor mass and evolution.

Here, we present optical spectra from four of these radio candidates as listed in Table 1 and shown in Figs.~1 through 4. For each, we give its radio name, position (R.A., Dec. (J2000)) at 3 cm (8.64 GHz), difference of radio and optical positions, radio spectral index (including error), radio diameter and optical counterpart PNe names. These known optical PNe have been found in numerous  surveys and were reconfirmed by Jacoby \& De Marco (2002).  They are listed in their Table 4, from which we use the abbreviation {\bf JD} followed by the object's number (i.e. JD04, JD10, JD26 and JD28). From Table 1, we also note that our spectral indices support the general belief that PNe are weak thermal radio emitters.

Spectral observations  were conducted January 2008, using the 1.9-meter telescope and Cassegrain spectrograph at the South African Astronomical Observatory (SAAO) in Sutherland.  We used grating number 7 (300 lpmm) to obtain spectra between 3500 and 6200 \AA\ having a  resolution of 5 \AA. For these, the slit size was $1.5\arcsec \times 1.5\arcmin$ with a spatial resolution of 0.74\arcsec\ pixel$^{-1}$.  Exposure times were limited to 800~s with a positional accuracy of $< 1$\arcsec.

Data reduction included bias subtraction and flat-field correction using the {\sc iraf} software package. Extraction (task `extractor'), including background sky subtraction, of data allowed the creation of one-dimensional spectra, wavelength calibrated using standard lines from a CuAr arc. Flux calibration was applied using the spectrometric standard star EG 21. Observing conditions were not photometric, seeing was limited to an arcsecond at best but varied throughout the evening.

\section{3. SPECTRAL ANALYSIS}\label{analysis}

We used {\sc iraf}'s task {\sc splot} to view and analyze our spectra. Only fluxes from spectral lines visually distinct from the baseline rms ($\sim 1.0 \times 10^{-15}$ ergs cm$^{-2}$ s$^{-1}$ \AA$^{-1}$) were selected for inclusion in Table 2. For each PNe, Table 2 lists the relative flux densities (using EG21) and 90\% confidence intervals of common lines within our spectral range.  These lines include the \OII 3727\AA\ doublet, \NeIII 3869\AA, H$\beta \lambda4861$\AA\ and \OIII 4363, 4959, 5007\AA.  All values are shown at their rest wavelengths.

We estimate extinction using Balmer emission lines H$\beta 4862$\AA, H$\gamma  4340$\AA\ and H$\delta  4102$\AA.  Noting that our spectral range does not include H$\alpha  6563$\AA, we
calculate Balmer decrements as H$\gamma$/H$\beta$ and H$\delta$/H$\beta$.  Characteristic
extinction curves are presented by Osterbrock \& Ferland (2006) and
expected intrinsic Balmer decrements are based on Case B recombination\footnote{
An approximation   characterized by large optical depth,
where every Lyman-line photon is scattered many times and is eventually
converted into lower-series photons.} Table 3 lists individual
values for E(B--V) based on these decrements. We cannot 
accurately measure the  H$\delta4102$\AA\  line for JD10, because of  baseline noise (Fig. 6).  

Of the available  decrements, the calculated extinction for each PN are similar.  These values 
are less than zero for JD04, JD10 and JD26, for which we were able to determine electron 
temperature ($T_{e}$) without any correction for extinction.  We 
cannot determine $T_{e}$ for JD28, since the \OIII4363\AA\ line is too
 weak\footnote{Extinction correction for JD28 is left to the reader.}.

The 90\% confidence errors reported in Table 2 are based on line measurement;  we
do not account for absolute photometric errors. Line measurement errors
were calculated using Monte-Carlo simulation techniques found in the task {\sc splot}
for a sample number of 100 and measured rms sensitivity. This 1$\sigma$ value was
multiplied by 1.64 to estimate a 90\% confidence interval for each flux density.





}

\end{multicols}

\newpage

{{\bf Table 1.} Radio PN Candidates in the SMC.  $\Delta$P represents the distance between radio and optical
positions.  JD refers to Table 4 found in Jacoby \& De Marco (2002), SMP
to Sanduleak et al. (1978), J to Jacoby (1980) and N to Henize (1956).
Spectral index ($\alpha$) is defined here as $S_{\nu} \propto \nu^{\alpha}$, where $S_{\nu}$ is
the flux density at frequency $\nu$.  }
\vskip2mm
\centerline{\begin{tabular}{|c|c|c|c|c|c|c|c|}
\hline
No. & ATCA Radio &R.A. & Dec. & $\Delta$P & $\alpha\pm\Delta\alpha$& Radio Dia. & Optical  \\
& Source Name& (J2000.0)&(J2000.0) &(arcsec) & &(arcsec/pc) & PN Name \\
\hline
1&J004336-730227 &00 43 36.54 &--73 02 27.1 &1\arcsec &$-0.3\pm0.3$&2\arcsec/0.6 &JD04, N9  \\
2&J004836-725802 &00 48 36.58 &--72 58 02.0 &2\arcsec &$0.0\pm0.1$ &2\arcsec/0.6 &JD10, SMP11, J8  \\
3&J005730-723224 &00 57 30.00 &--72 32 24.0 &2\arcsec &$0.0\pm0.1$ &2\arcsec/0.6 &JD26, N61  \\
4&J005842-722716 &00 58 42.89 &--72 27 16.5 &1\arcsec &$0.0\pm0.1$ &2\arcsec/0.6 &JD28, N68  \\
\hline
\end{tabular}}

\vskip.5cm

\begin{multicols}{2}
{{\ }


\centerline{\includegraphics[height=6.0cm,angle=270]{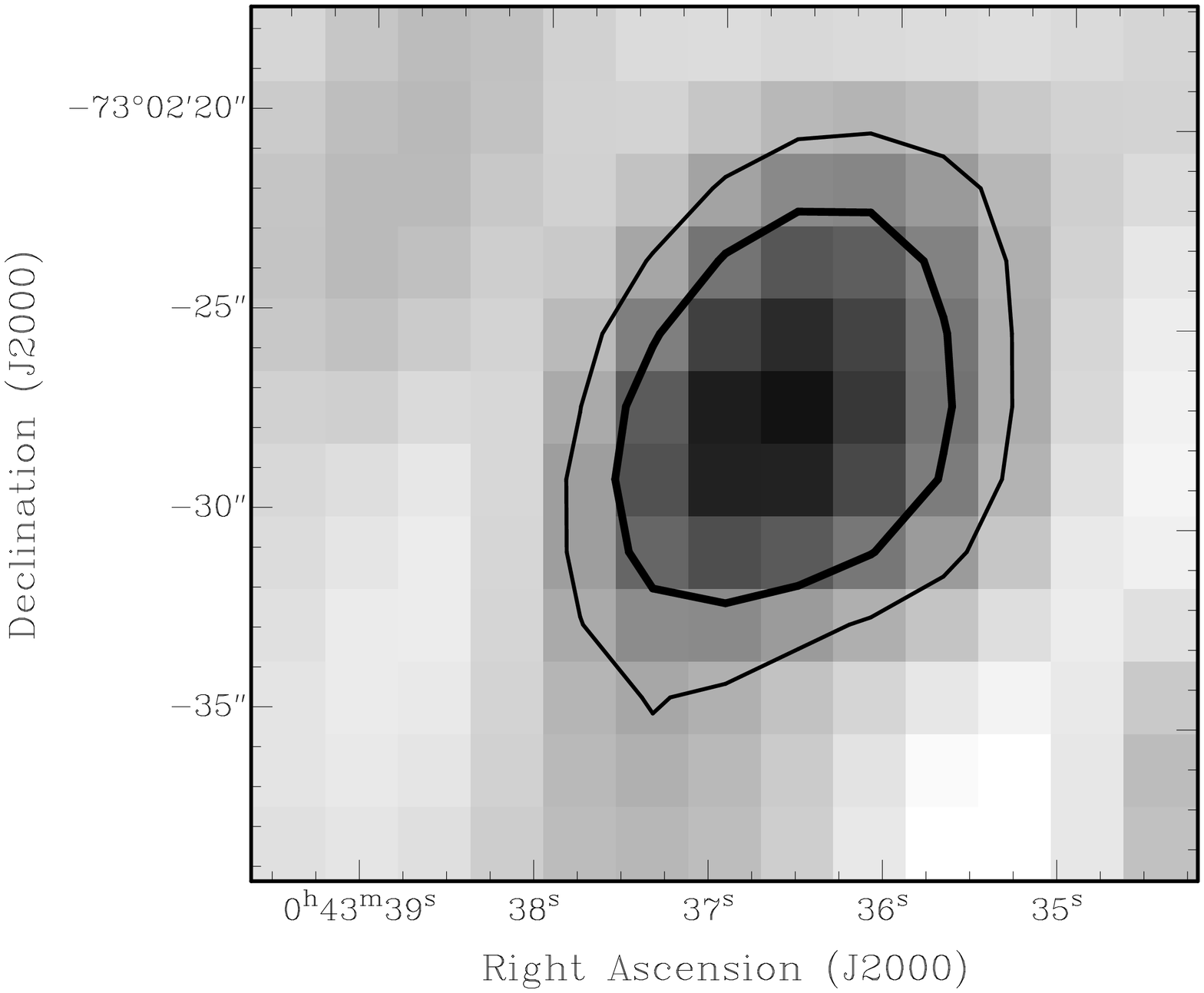}}


\figurecaption{1.}{20 cm image $\mathrm{(beam = 7\arcsec \times 8\arcsec)}$ of J004336-730227 with 20 cm contours at 0.3 and 0.6 mJy beam$^{-1}$.} }

{{\ }


\centerline{\includegraphics[height=6.0cm,angle=270]{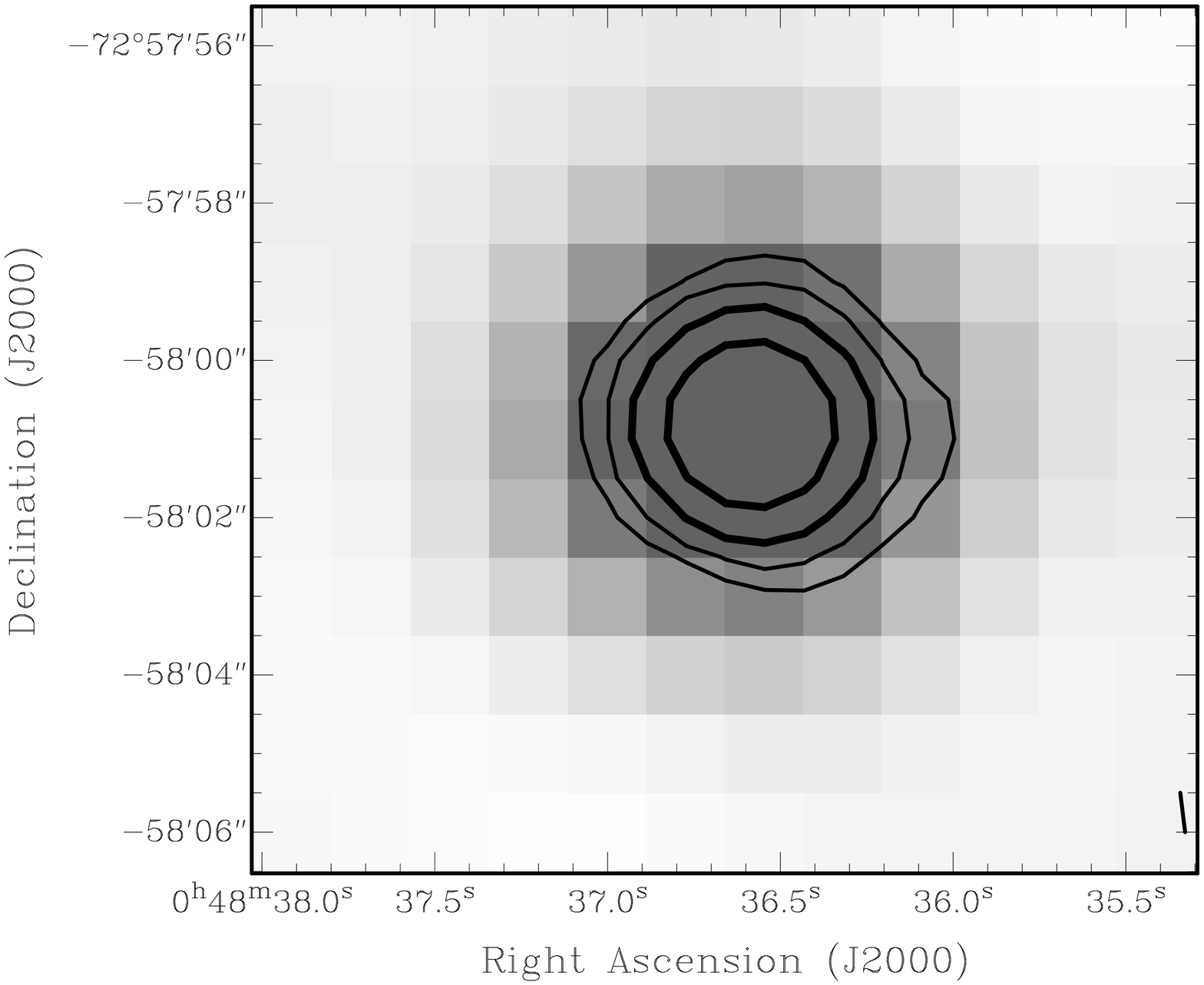}}


\figurecaption{2.}{6 cm image $\mathrm{(beam = 4\arcsec)}$ of J004836-725802 with 3 cm contours
$\mathrm{(beam = 2\arcsec)}$ at 0.15, 0.3, 0.6 and 1.2 mJy beam$^{-1}$.} }

\end{multicols}

\begin{multicols}{2}

{{\ }


\centerline{\includegraphics[height=6.0cm,angle=270]{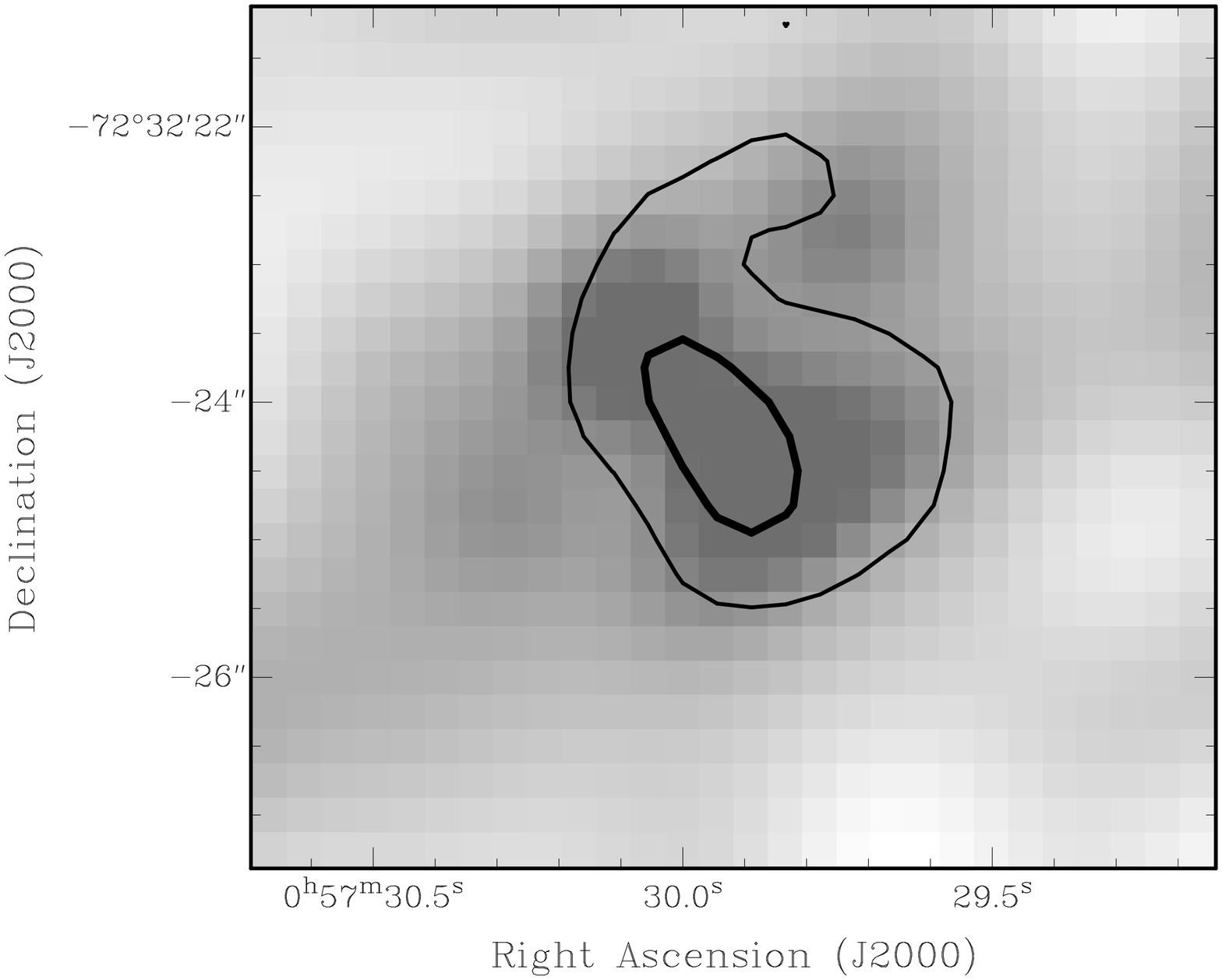}}


\figurecaption{3.}{6 cm image $\mathrm{(beam = 4\arcsec)}$ of J005730-723224 with 3 cm contours $\mathrm{(beam = 2\arcsec)}$ at 0.3 and 0.6 mJy beam$^{-1}$.} }

{{\ }


\centerline{\includegraphics[height=6.0cm,angle=270]{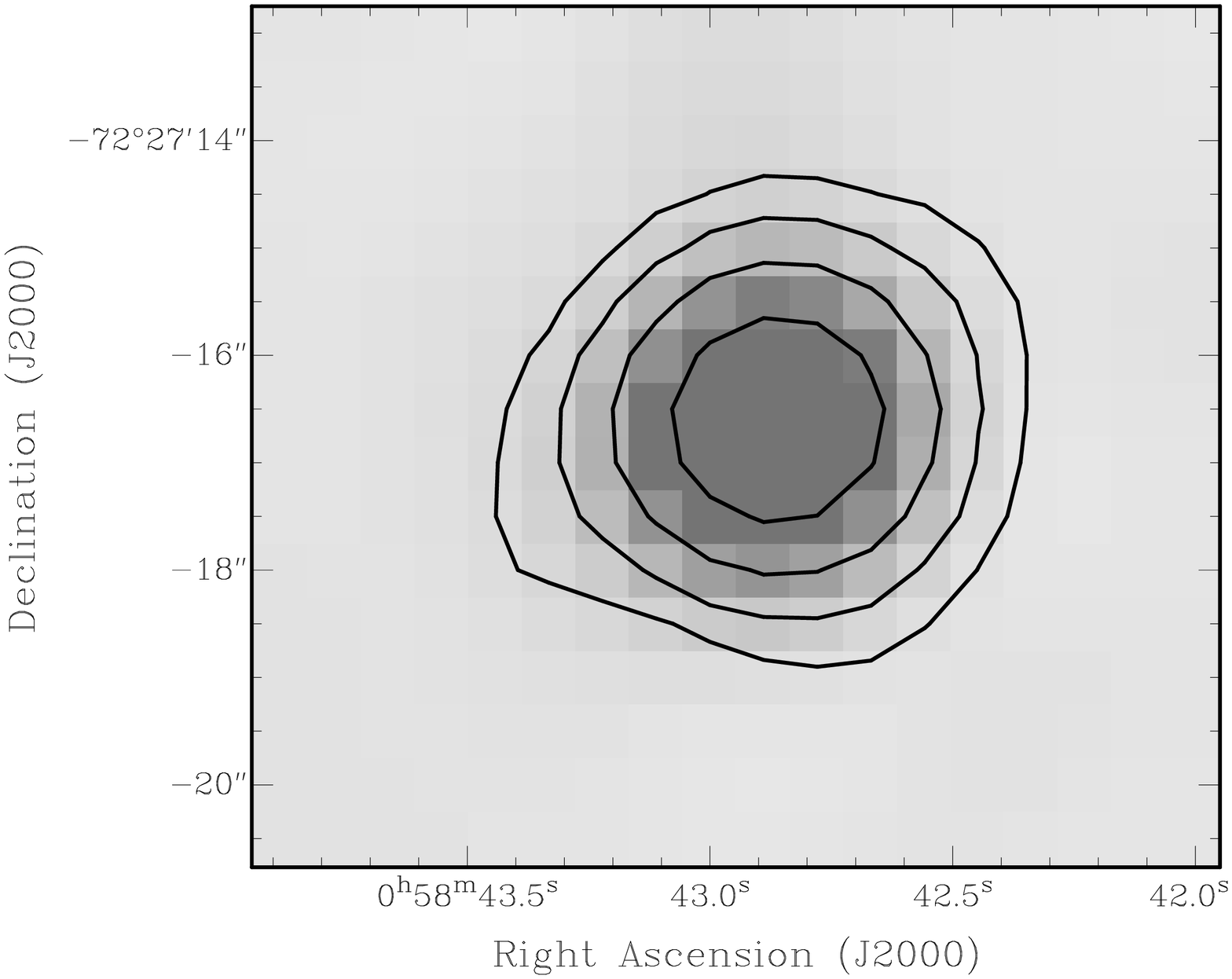}}


\figurecaption{4.}{6 cm image $\mathrm{(beam = 4\arcsec)}$ of J005842-722716 with 3 cm contours
$\mathrm{(beam = 2\arcsec)}$ at 0.3, 0.6, 1.2 and 2.4 mJy beam$^{-1}$.} }

\end{multicols}

\newpage


{{\bf Table 2.} Spectral Analysis of Known Optical PNe Co-Incident with Radio Sources.  Flux density is given in units of $10^{-14}$ ergs cm$^{-2}$ s$^{-1}$ and include 90 \% confidence intervals.  Electron temperature ($T_{e}$) and ionized mass ($M_{i}$) both
assume an electron density ($n_{e}$) of  $10^{3}$ cm$^{-3}$.  There is no correction for extinction.   }
\vskip2mm
{\small
\centerline{\begin{tabular}{|c|c|c|c|c|c|c|c|c|c|}
\hline
Name & \OII        &\NeIII                &\OIII                     & H$\beta$         & \OIII               &\OIII         &$T_{e}$&S$_{4.8\mathrm{GHz}}$& $M_{i}$ \\
            & 3727\AA            & 3869\AA         &4363\AA            &4861\AA           &4959\AA       &5007\AA &(K)         &(mJy)& ($M_{\odot}$)\\
\hline
JD04  &$1144.0\pm1.0$&$82.6\pm1.0$ &$11.9\pm1.0$ &$186.6\pm1.0$&$196.0\pm1.0$&$582.0\pm1.0$ &15369&2.6&2.6  \\
JD10  &$82.5\pm1.0$     &---                     &$2.9\pm1.5$    &$31.0\pm1.0$   &$34.5\pm1.0$ &$100.5\pm1.0$ &18412 &2.6&2.6 \\
JD26  &$659.1\pm0.9$  &$13.7\pm1.0$ &$19.5\pm0.7$ &$96.5\pm0.9$    &$55.2\pm1.0$ &$163.2\pm1.0$ &95759&3.0&3.0  \\
JD28  &$6.3\pm0.7$       &$6.0\pm1.1$   & ---                     &$21.7\pm1.0  $  &$33.7\pm1.0$ &$106.4\pm1.0$ &--- &4.5& 4.6\\
\hline
\end{tabular}}
}
\vskip.5cm

\begin{multicols}{2}
{{\ }


\centerline{\includegraphics[height=6.0cm]{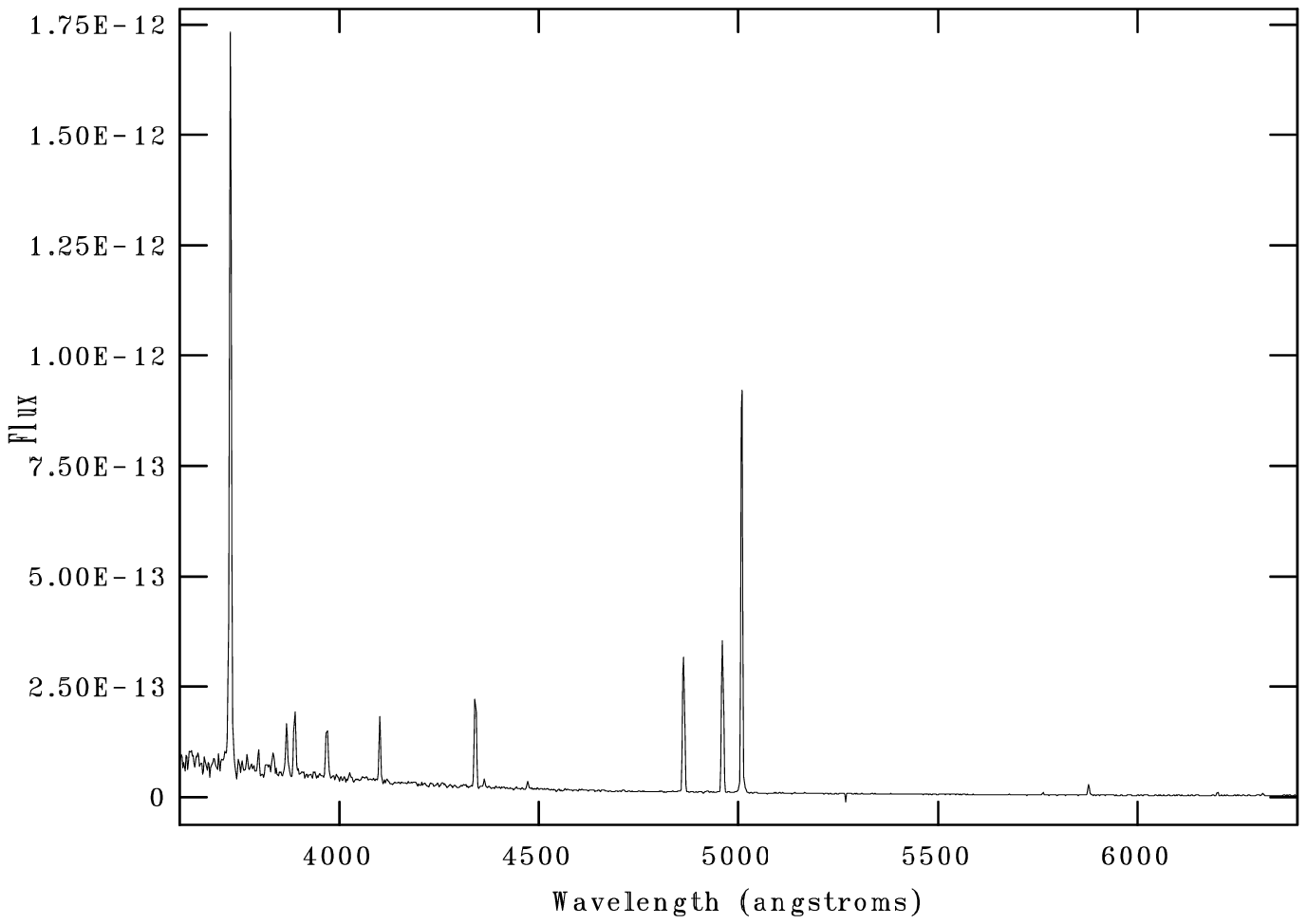}}


\figurecaption{5.}{Optical spectrum of JD04 coincident with J004336-730227.} }

{{\ }


\centerline{\includegraphics[height=6.0cm]{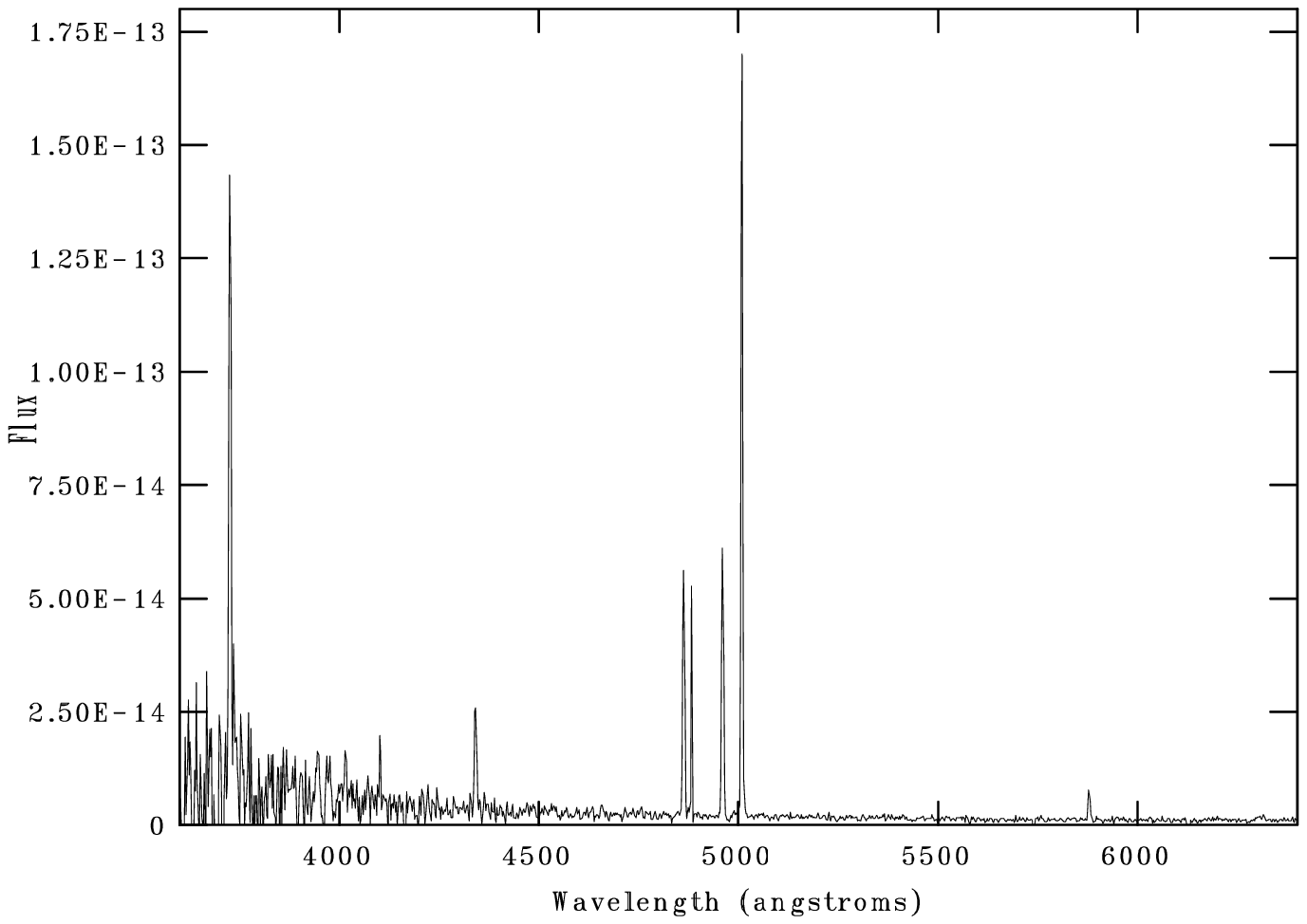}}


\figurecaption{6.}{Optical spectrum of JD10 coincident with J004836-725802.} }

\end{multicols}

\begin{multicols}{2}

{{\ }


\centerline{\includegraphics[height=6.0cm]{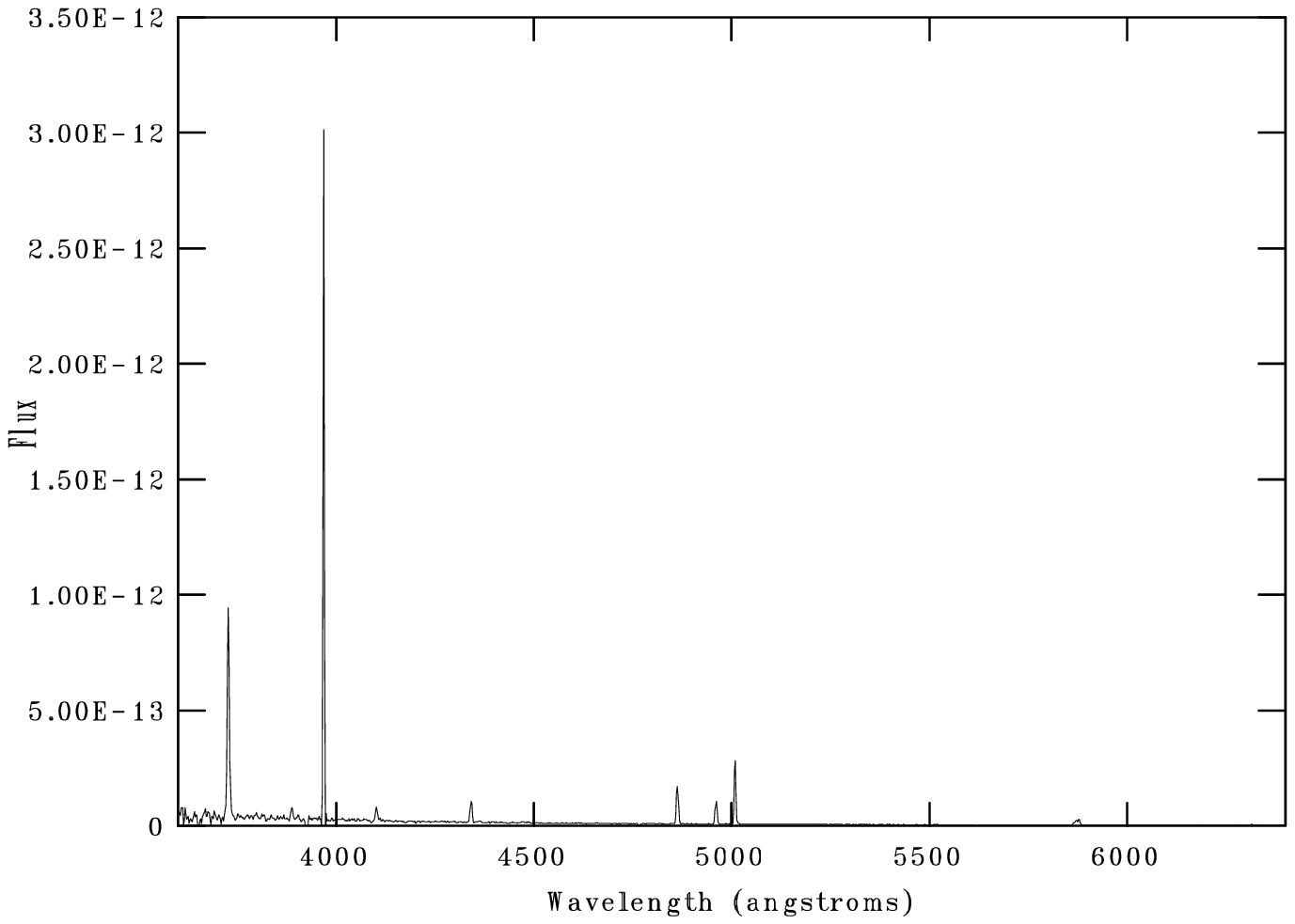}}


\figurecaption{7.}{Optical spectrum of JD26 coincident with J005730-723224.} }

{{\ }


\centerline{\includegraphics[height=6.0cm]{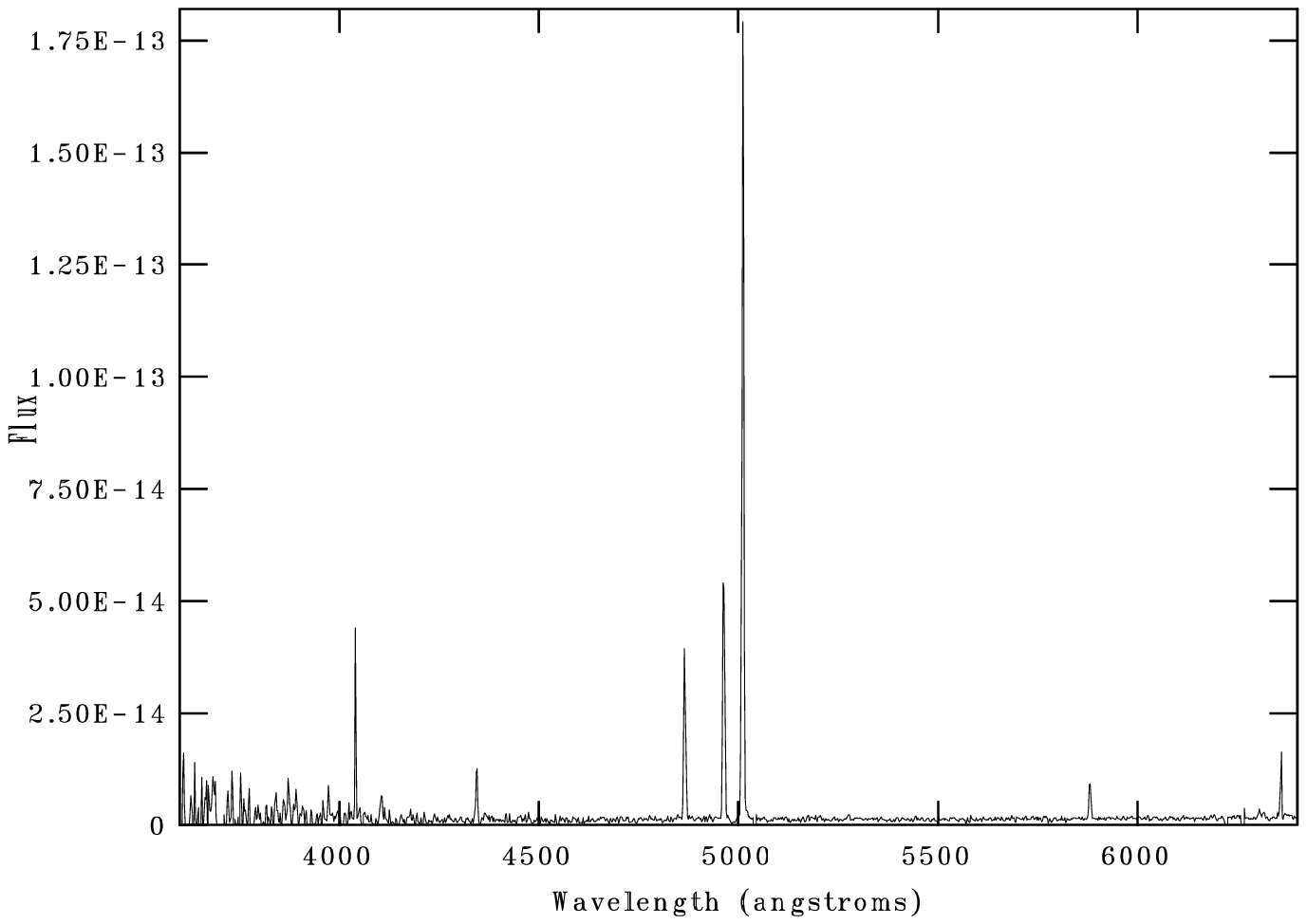}}


\figurecaption{8.}{Optical spectrum of JD28 coincident with J005842-722716.} }

\end{multicols}







\begin{multicols}{2}
{




Ions including \OIII\ and \NII\ have energy-level structures that
produce emission lines from different excitation energies.
The relative rates of excitation depend very strongly on temperature and may
be used to measure $T_{e}$.  The only practical probe for the
measurement of electron temperature in our case is the \OIII\ forbidden line
ratio\footnote{Defined as ${\lambda 4959 + \lambda 5007} \over \lambda 4363$.}.
We list electron temperatures in Table 2 for each of our nebulae, based on this
\OIII\ ratio.  

Since our spectral range excluded  the
\SII$6717, 6731$\AA\ lines needed to calculate electron densities, we 
assume a $n_{e}$ of $10^{3}$ cm$^{-3}$ for this selected radio sample.
We base this rough value on   PNe averages  found in the literature 
(see Stanghellini \& Kaler 1989) and estimate
$T_{e}$ using the Space Telescope Science Data Analysis System task
{\sc nebular.temden}, based on an five-level atom approximation from De Robertis
et al. (1987).

If one assumes these PNe have a relatively symmetric uniform density,
it may be possible to crudely estimate their ionized Mass ($M_{i}$):

\begin{equation}
M_{i} = 282\mathrm{(}D_{\mathrm{kpc}}\mathrm{)}^{2}F_{5}\mathrm{(}n_{e}\mathrm{)}^{-1}M_{\odot},
\end{equation}

\noindent where $D_{\mathrm{kpc}}$ is distance (kpc), $F_{5}$ is radio flux density at $\sim 5$ GHz (Jy) and $n_{e}$ represents assumed electron density (cm$^{-3}$) (Kwok 2000).
Table 2 lists an ionized mass estimate for each PN based on the given flux densities
at 4.8~GHz. This implies that our selected radio bright objects have much higher
nebular ionized mass than is typically expected.

\vskip.5cm

{{\bf Table 3.} Extinction Values for Balmer Decrements. }
\vskip2mm
\centerline{\begin{tabular}{|c|c|c|c|c|}
\hline
Name&H$\gamma$/H$\beta$ & E(B--V) & H$\delta$/H$\beta$ &E(B--V)\\
\hline
JD04&0.7 &--0.8 &0.4 &-0.8 \\
JD10&0.5 &--0.2 & --& --\\
JD26&0.7 &--0.8 &0.5 &-0.9 \\
JD28&0.4 & 0.6&0.2 &0.6 \\
\hline
\end{tabular}}

\vskip1.0cm

Figs. 5 through 8 show one-dimentional spectra for each PNe listed in Table 2.  These spectra have fairly typically emission lines with a few exceptions.  For example, Fig. 6 shows that JD10 has a predominant \FeIII 4881\AA\ line.  This is unusual since the abundance of iron in PNe is scarcely studied, due to the relative faintness of observed iron emission lines (Perinotto et al. 1999).  Therefore, such a strong line warrants further verification and study.

We also note in Fig. 7 (JD26), the spectrum has a very strong emission line at 3968\AA\ that
may represent a blend of \NeIII\ and H$\epsilon$. In Fig. 8 (JD28), \HeI\ lines are
prominent at 4026, 5876\AA\ in addition to a strong Balmer line
at H$\gamma 4341$\AA.


\section{4. CONCLUDING REMARKS AND SUMMARY}\label{summary}

Most observed PNe have nebular masses of only 0.3 $M_\odot$, although the main-sequence mass of PNe progenitors can be as high as 8 $M_\odot$ (Kwok 1994). We believe the PNe we are studying may represent a predicted ``missing-mass link'', belonging originally to a system possessing up to an 8 $M_\odot$ central star.

High rates of mass loss that continue for an extended fraction of a AGB's lifetime can allow a significant fraction of the star's mass to be accumulated.  This may result in the formation of a circumstellar envelope (CSE). If the transition from the AGB to PN stage is short, then such CSEs could have a significant effect. Perhaps, our radio observations select for high mass PNe, since the quantity of ionized mass present appears directly related to radio flux density.

Our combined observations suggest that a population of PNe in the SMC have bright radio counterparts within 2\arcsec\ of each other. These PNe have fairly typical optical spectra with the expected emission lines.  Nebular electron temperatures are also within the expected range assuming an average density of $10^3$ cm$^{-3}$. Given values of radio flux density at $\sim5$ GHz, we suggest that the ionized nebular mass of these PNe may be 2.6 $M_\odot$ or greater.


\acknowledgements{This paper uses observations made from the
South African Astronomical Observatory (SAAO).   Travel to the SAAO was funded by
Australian Government, AINSTO AMRFP proposal number 07/08-O-11.
We also used the {\sc karma} software package developed by the ATNF.  The
Australia Telescope Compact Array is part of the Australia Telescope
which is funded by the Commonwealth of Australia for operation as a
National Facility managed by CSIRO.  {\sc iraf} is distributed
by the National Optical Astronomy Observatories, which are
operated by the Association of Universities for
Research in Astronomy, Inc., under cooperative
agreement with the National Science Foundation.}


\references

De Robertis, M. M., Dufour, R., Hunt, R.: 1987, \journal{JRASC}, \vol{81}, 195.


Filipovi\'c, M. D., Bohlsen, T., Reid, W., Staveley-Smith, L., Jones, P. A., Nohejl, K., Goldstein, G.: 2002, \journal{MNRAS}, \vol{335}, 1085.

Henize, K. G.: 1956, \journal{ApJS}, \vol{2}, 315.

Hilditch, R. W., Howarth, I. D., Harries, T. J.: 2005, \journal{MNRAS}, \vol{357}, 304.

Jacoby, G. H.: 1980, \journal{ApJS}, \vol{42}, 1.

Jacoby, G. H., De Marco, O.: 2002, \journal{AJ}, \vol{123}, 269.

Kwok, S.: 1994, \journal{PASP}, \vol{106}, 344.

Kwok, S.: 2000, "The Origin and Evolution of Planetary Nebulae", Cambridge University Press,
Cambridge, p. 51.

Kwok, S.: 2005, \journal{JKAS}, \vol{38}, 271.

Osterbrock, D. E., Ferland, G. J.: 2006, "Astrophysics of Gaseous Nebulae 
and Active Galactic Nuclei (second edition)".  University Science Books, Sausalito, p. 180.

Payne, J. L., Filipovi\'c, M. D., Reid, W., Jones, P. A., Staveley-Smith, L, White, G. L.:2004, \journal{MNRAS}, \vol{355}, 44.

Perinotto, M., Bencini, C. G., Pasquali, A., Manchado, Rodriguez Espinosa, J. M., Stanga, R.: 1999,
\journal{A\&A}, \vol{347}, 967.

Reid, W. A., Parker, Q. A.: 2006, \journal{MNRAS}, \vol{373}, 521.

Sanduleak, N., MacConnell, D. J., Philip, A. G. Davis: 1978, \journal{PASP}, \vol{90}, 621.

Stanghellini, L., Kaler, J. B.: 1989, \journal{ApJ}, \vol{343}, 811.





\endreferences

}
\end{multicols}

\vfill\eject

{\ }



\naslov{Optiqki spektri radio planetarnih maglina u Malom Magelanovom Oblaku}


\authors{J. L. Payne$^{1}$, M. D. Filipovi\'c$^{2}$, E. J. Crawford$^{2}$, A.Y.~De~Horta$^{2}$, G. L. White$^{1}$ and F.H.~Stootman$^{2}$}

\vskip3mm


\address{$^1$Centre for Astronomy, James Cook University\break Townsville QLD, 4811, Australia}

\address{$^2$University of Western Sydney, Locked Bag 1797,\break Penrith South DC, NSW 1797, Australia}

\vskip.7cm




\centerline{\rit }

\vskip.7cm

\begin{multicols}{2}
{

\rrm U ovoj studiji predstav{lj}amo preliminarne rezultate spektralne analize 4 radio kandidata za planetarnu maglinu u Malom Magelanovom Oblaku (MMO). Optiqka posmatra{nj}a prikazana ovde, su ura{dj}ena sa Radklif 1.9-m teleskopom (Saderlend, Ju{\zz}na Afrika). Ove 4 radio planetarne magline originalno su otkrivene sa naxih ranijih radio pregleda MMO naprav{lj}enih sa Australijanskim Teleskopom Kompaktnog Po{lj}a (ATCA) na 20 i 13 cm, a tako{dj}e su potvr{dj}ene dodatnim posmatra{nj}ima visoke rezolucije na 6 i 3 cm (4\arcsec/2\arcsec). Optiqke i radio pozicije ova 4 kandidata za radio planetarnu maglinu u MMO su uda{lj}eni ispod 2\arcsec\ i najverovatnije predstav{lj}aju pod-populaciju planetarnih maglina izra{\zz}enih kao veoma sjajni radio objekti. Nebularne jonizovane mase ovih objekata su proraqunate na oko 2.6~$M_\odot$ ili vixe, xto da{lj}e podr{\zz}ava postoja{nj}e planetarnih maglina sa progenitronskim centralnim zvezdama masa do 8 $M_\odot$.
}
\end{multicols}

\end{document}